\documentclass[a4paper,12pt]{article}
\usepackage[T2A]{fontenc} 
\usepackage[russian, english]{babel} 
\usepackage[dvips]{graphicx} 
\makeindex 
\usepackage{epsfig} 
\usepackage[dvips]{texdraw} \input{txdtools}
\usepackage{picinpar} 
\usepackage{amsmath}
\usepackage{indentfirst} 

\begin{document}
\begin{center}

{\bf \large Upper Limits on Electric and Weak Dipole Moments \\ of
$\tau$-Lepton and Heavy Quarks from $e^+e^-$ Annihilation}

\vspace{0.5cm} 

A.~E.~Blinov and A.~S.~Rudenko

\vspace{0.5cm}

Budker Institute of Nuclear Physics,\\ 630090 Novosibirsk, Russia,\\ and Novosibirsk
State University

\end{center}

\begin{abstract}
The total cross-sections measured at LEP for $e^+e^-$ annihilation into
$\tau^+\tau^-$, $c\bar{c}$ and $b\bar{b}$ at $2E \simeq 200$ GeV are used to derive
the upper limits\, $3\cdot10^{-17}$,\, $5\cdot10^{-17}$,\, $2\cdot10^{-17}$\,
$e\cdot$cm for the electric dipole moments and\, $4\cdot10^{-17}$,\,
$7\cdot10^{-17}$,\, $2.5\cdot10^{-17}$\, $e\cdot$cm for the weak dipole moments of
the\, $\tau$-lepton,\, $c$-, and\, $b$-quarks, respectively. Some of the existing
limits on these moments are improved and for the $b$-quark the improvement is rather
significant.
\end{abstract}

\subsection*{1. Theory}

The existence of the electric dipole moment (EDM) $d$ and weak dipole moment (WDM)
$d^w$ would imply $CP$ violation. Since the expected values of $d$ and $d^w$ in the
Standard Model (SM) are extremely small the measurement of significantly larger
values would be evidence for physics beyond the SM.

\begin{figure}[h]
\center
\begin{tabular}{c c}
\includegraphics[scale=1.2]{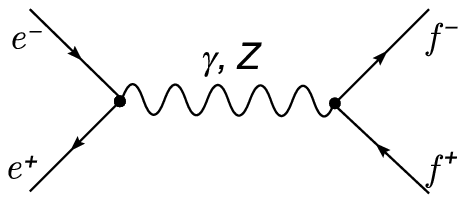} &
\includegraphics[scale=1.2]{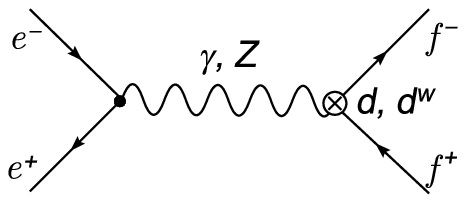} \\
(a) & (b)
\end{tabular}
\caption {(a) the diagrams with the regular vertices; \newline \hspace*{1.67 cm} (b)
the diagrams with the dipole moments vertices} \label{fig:1}
\end{figure}

In this paper we consider the reaction of the high-energy electron-positron
annihilation into fermions supposing that the final particles have the EDM $d$ and
WDM $d^w$ (see the corresponding Feynman graphs at tree level in Fig.~\ref{fig:1}).

The final fermions could be both leptons and quarks (if the quark-antiquark pair
production occurs far above the threshold and interaction between them may be
neglected).

The effective Lagrangian describing interaction of the EDM with the electromagnetic
field and the WDM with the $Z$-boson field is

\begin{equation}
L_{eff}=-\frac{1}{2}\bar\psi\gamma_5\sigma^{\mu\nu}\psi (d
F_{\mu\nu}+d^w F^Z_{\mu\nu}),
\end{equation}
where $\gamma_5=-i\gamma_0\gamma_1\gamma_2\gamma_3=\left(%
\begin{array}{cc}
  0 & -I \\
  -I & 0 \\
\end{array}%
\right)$,
$\sigma^{\mu\nu}=\frac{1}{2}\left(\gamma^\mu\gamma^\nu-\gamma^\nu\gamma^\mu\right)$.

The EDM and WDM are supposed to have no ima\-gi\-nary parts, since the latter would
violate not only $CPT$ invariance, but also Hermitian character of the Lagrangian.

Fermions in the reaction $e^+e^- \to f\bar f$ are produced in triplet states if the
production mechanism is regular: $^3S_1$ and $^3D_1$ if the vertex is a vector
$\gamma_\mu$, or $^3P_1$ if the vertex is an axial one $\gamma_5\gamma_\mu$.
However, if produced via the $CP$-odd electric or weak dipole moment vertex $d
\gamma_5\sigma_{\mu\nu}q_\nu$ or $d^w \gamma_5\sigma_{\mu\nu}q_\nu$, the fermions
will be in the singlet state $^1P_1$. Therefore, if polarization of the final
particles is not taken into account the dipole moment vertices do not interfere with
the regular ones, and hence their contribution to the cross-section is of second
order in the dipole moments.

Considering the reaction in the centre-of-mass system assuming unpolarized electron
and positron beams, neglecting the electron mass and summing over pola\-ri\-zation
of the final particles we obtain the following expressions of the squared matrix
elements $M^2_d$, $M^2_{d^w}$ and the interference term $M^2_{d d^w}$:

\begin{equation}
M^2_d=4e^2d^2E^2\left(1-\frac{m_f^2}{E^2}\right)\sin^2\theta,
\end{equation}

\begin{equation}
M^2_{d^w}=\frac{16 e^2 (d^w)^2}{\sin^2\theta_W\cos^2\theta_W}
\frac{E^4\left(E^2-m^2_f\right)}{\left(4E^2-m_Z^2\right)^2+m_Z^2\Gamma_Z^2}\left(V_e^2+A_e^2\right)\sin^2\theta,
\end{equation}

\begin{equation}
M^2_{d d^w}=-\frac{16 e^2 d
d^w}{\sin\theta_W\cos\theta_W}\frac{E^2\left(E^2-m^2_f\right)\left(4E^2-m_Z^2\right)}{\left(4E^2-m_Z^2\right)^2+m_Z^2\Gamma_Z^2}V_e\sin^2\theta,
\end{equation}

\noindent where $E$ is the beam energy, $V_e=-\frac{1}{2}+2\sin^2 \theta_W$,
$A_e=-\frac{1}{2}$.

The total cross-section of the process $e^+e^-\rightarrow f\bar f$ looks as follows:

\begin{equation}
\sigma=\sigma_{SM}+
\frac{1}{32\pi}\frac{1}{4E^2}\sqrt{1-\frac{m_f^2}{E^2}}N_c\int\left(M^2_d+M^2_{d^w}+M^2_{d
d^w}\right) d(\cos\theta),
\end{equation}

\noindent where $N_c=1$ if the final particles are leptons and $N_c=3$ for quarks.

The interference term $M^2_{d d^w}$ is numerically suppressed due to smallness of
$V_e$.

\subsection*{2. Analysis of experimental data}

Since the relative contribution of the $M^2_d$ and $M^2_{d^w}$ terms grows with
energy, the best limits can be obtained from the highest energy experiments.
Therefore, we use the LEP-II measurements of the two-fermion final states combined
by LEPEWWG \cite{LEPEWWG} in order to set the limits. The integrated luminosity
amounts to about 670 $pb^{-1}$ per experiment collected at centre-of-mass energies
between 183 and 207 GeV.

The measured values and the SM predictions of $e^+e^- \to \tau^+ \tau^-$
cross-sections were obtained from Table~3.2 of Ref.~\cite{LEPEWWG}. The measured
values and the SM predictions of $e^+e^-\to c\bar c$ and  $e^+e^-\to b\bar b$
cross-sections were obtained as a combination of $e^+e^-\to q\bar q$ ones (Table~3.2
of Ref.~\cite{LEPEWWG}) and $R_c$- and $R_b$-fractions of the $c\bar c$\, and $b\bar
b$ among the $q\bar q$ final states (Tables~3.9, 3.10 of Ref.~\cite{LEPEWWG}). The
errors of the $c\bar c$\, and $b\bar b$ cross-sections are dominated by the
uncertainties of $R_c$ and $R_b$, respectively.

The measured cross sections coincide within errors with the SM predictions. The
$\chi^2$ fits of residuals by the second term in Formula (5) with non-zero $d$- or
$d_w$-terms yield the 95\% CL limits $3\cdot10^{-17}$, $5\cdot10^{-17}$,
$2\cdot10^{-17}$ $e\cdot$cm for the electric and $4\cdot10^{-17}$, $7\cdot10^{-17}$,
$2.5\cdot10^{-17}$ $e\cdot$cm for the weak dipole moments of the $\tau$-lepton,
$c$-, and $b$-quarks, respectively.

\subsection*{3. Discussion}

\renewcommand{\arraystretch}{1.1}

\begin{table*}[t]
\begin{center}
\caption{Comparison of our results with the previous ones} \label{table:1}
\renewcommand{\tabcolsep}{2pc} 
\renewcommand{\arraystretch}{1.1} 
\begin{tabular}{|c|c|} \hline
Previous limits ($e\cdot$cm) & Our limits ($e\cdot$cm) \\ \hline
$d_\tau<1.4 \cdot 10^{-16}$ \cite{del Aguila}, & \\
$-2.2<\mathrm{Re}(d_\tau)<4.5$ ($10^{-17}$) and & \\ $-2.5<\mathrm{Im}(d_\tau)<0.8$
($10^{-17}$) \cite{Inami}, & \\ $d_\tau <1.1 \cdot 10^{-17}$ \cite{Escribano} &
$d_\tau<3\cdot10^{-17}$ \\ \hline
$d_c<8.9\cdot10^{-17}$ \cite{Escribano2} & $d_c<5\cdot10^{-17}$ \\
\hline
$d_b<8.9\cdot10^{-17}$ \cite{Escribano2} & $d_b<2\cdot10^{-17}$ \\
\hline $|\mathrm{Re}(d^w_\tau)|<5.0\cdot 10^{-18}$ and & \\
$|\mathrm{Im}(d^w_\tau)|<1.1\cdot 10^{-17}$ \cite{ALEPH}, & \\
$-3.56<\mathrm{Re}(d^w_\tau)<2.26$ ($10^{-18}$) and & \\
$-0.69<\mathrm{Im}(d^w_\tau)<0.77$ ($10^{-17}$) \cite{Lohmann} & $d^w_\tau<4\cdot10^{-17}$ \\
\hline $d^w_c<5.7\cdot10^{-17}$ \cite{Rizzo} & $d^w_c<7\cdot10^{-17}$ \\ \hline
$d^w_b<6.0\cdot10^{-16}$ \cite{Rizzo} & $d^w_b<2.5\cdot10^{-17}$
\\ \hline
\end{tabular}
\end{center}
\end{table*}

Now we briefly compare our results with other ones. For the $\tau$-lepton the upper
limit\linebreak $d_\tau< 1.4 \cdot 10^{-16}$ $e\cdot$cm was obtained from the
ana\-ly\-sis of angular distribution of the tau pairs in the $e^+e^- \to \tau^+
\tau^-$ reaction at $2E \simeq 35$ GeV \cite{del Aguila}. The best direct
experimental restrictions on the $\tau$ EDM are $-2.2<\mathrm{Re}(d_\tau)<4.5$
($10^{-17}$ $e\cdot$cm) and \linebreak $-2.5<\mathrm{Im}(d_\tau)<0.8$ ($10^{-17}$
$e\cdot$cm) \cite{Inami}. The result was obtained in the $e^+e^- \to \tau^+ \tau^-$
reaction at $2E \simeq 10$ GeV using a technique which takes into account $\tau$
polarization. The even better upper limit $d_\tau <1.1 \cdot 10^{-17}$ $e\cdot$cm
was obtained from the partial width $\Gamma(Z \to \tau^+ \tau^-)$ measured at LEP in
Z peak \cite{Escribano}. However, a model dependent relationship between the weak
and electric dipole moments is used. Quite recently upper limit $d_\tau < (1 \pm 1)
\cdot 10^{-16}$ $e\cdot$cm at momenta about $m_\tau \sim 1$ -- 2 GeV have been
derived from the precision measurements of the electron EDM \cite{Grozin}.

For the $c$- and $b$-quarks the limits $d_c<8.9\cdot10^{-17}$ $e\cdot$cm and
$d_b<8.9\cdot10^{-17}$ $e\cdot$cm were obtained from measurements at the $Z$ peak
\cite{Escribano2} with model assumptions similar to Ref.~\cite{Escribano}.

For the $\tau$-lepton the upper limits $|\mathrm{Re}(d^w_\tau)|<5.0\cdot 10^{-18}$
$e\cdot$cm, $|\mathrm{Im}(d^w_\tau)|<1.1\cdot 10^{-17}$ $e\cdot$cm were obtained
from the reaction $e^+e^- \to \tau^+ \tau^-$ at ener\-gies near the $Z$ resonance
\cite{ALEPH}. The results combining ALEPH, DELPHI and OPAL data lead to
$-3.56<\mathrm{Re}(d^w_\tau)<2.26$ ($10^{-18}$ $e\cdot$cm) and
$-0.69<\mathrm{Im}(d^w_\tau)<0.77$ ($10^{-17}$ $e\cdot$cm) \cite{Lohmann}.

The bounds on $d^w_c$ and $d^w_b$ were obtained from the decays $Z \to c \bar c$, $Z
\to b \bar b$ and they are $d^w_c<5.7\cdot10^{-17}$ $e\cdot$cm,
$d^w_b<6.0\cdot10^{-16}$ $e\cdot$cm \cite{Rizzo}.

Thus, the upper limits obtained from the contribution of the EDM to the total cross
section of $f\bar f$-production at LEP-II energy are similar to the best direct
experimental limits for $d_\tau$ and are somewhat tighter for $d_c$ and $d_b$.

Our upper limit on the $\tau$ WDM is worse than ones in Refs.~\cite{ALEPH,Lohmann}
because the analysis therein used a technique which takes $\tau$ polarization into
account. For the $c$ quark our bound on the WDM is similar to and for the $b$ is
better than ones in Ref.~\cite{Rizzo}.

For convenience we present a comparison of our results with the previous ones in
Table~\ref{table:1}.

\subsection*{Acknowledgements}

We would like to thank A.~E.~Bondar, V.~E.~Blinov and I.~B.~Khriplovich for many
useful discussions and suggestions. We are also grateful to I.~B.~Khriplovich and
A.~V.~Grabov\-sky for critical reading of the article.

\renewcommand{\bibname}

\end{document}